\documentstyle[12pt,aasms4]{article}

\def\gsim{\;\rlap{\lower 2.5pt
\hbox{$\sim$}}\raise 1.5pt\hbox{$>$}\;}
\def\lsim{\;\rlap{\lower 2.5pt
   \hbox{$\sim$}}\raise 1.5pt\hbox{$<$}\;}

\def\ge{\;\rlap{\lower 2.5pt
 \hbox{$-$}}\raise 1.5pt\hbox{$>$}\;}
\def\le{\;\rlap{\lower 2.5pt
   \hbox{$-$}}\raise 1.5pt\hbox{$<$}\;}
\def\HI{\ion{H}{1}~}
\def\HII{\ion{H}{2}~}

\newcommand\beq{\begin{equation}} 
\newcommand\eeq{\end{equation}}
\def\lya{Ly$\alpha$~}
\def\v{\vspace{-0.1in}}

\begin{document}

\Large \centerline{\bf A Size of $\sim10$Mpc for the Ionized Bubbles}
\Large \centerline{\bf at the End of Cosmic Reionization}

\normalsize
\author{\bf J. Stuart B. Wyithe$^\dagger$ and Abraham Loeb$^\star$}
\medskip
\noindent
$\dagger$ School of Physics, The University of Melbourne, Parkville, Vic
3010, Australia \\ $\star$ Astronomy Dept., Harvard University, 60 Garden
Street, Cambridge, MA 02138, USA\\

\vskip 0.2in 
\hrule 
\vskip 0.2in 

{\bf The first galaxies to appear in the universe at redshifts $z\ga 20$
created ionized bubbles in the intergalactic medium (IGM) of neutral
hydrogen (\HI) left over from the Big-Bang. It is thought that the ionized
bubbles grew with time, surrounded clusters of dwarf
galaxies\cite{BL1}$^,$\cite{FZH1} and eventually overlapped quickly
throughout the universe over a narrow redshift interval near $z\sim
6$. This event signaled the end of the reionization epoch when the universe
was a billion years old.  Measuring the hitherto unknown size distribution
of the bubbles at their final overlap phase is a focus of forthcoming
observational programs aimed at highly redshifted 21cm emission from atomic
hydrogen.  Here we show that the combined constraints of cosmic variance
and causality imply an observed bubble size at the end of the overlap epoch
of $\sim 10$ physical Mpc, and a scatter in the observed redshift of
overlap along different lines-of-sight of $\sim 0.15$. This scatter is
consistent with observational constraints from recent spectroscopic data on
the farthest known quasars. Our novel result implies that future radio
experiments should be tuned to a characteristic angular scale of $\sim
0.5^\circ$ and have a minimum frequency band-width of $\sim 8$ MHz for an
optimal detection of 21cm flux fluctuations near the end of reionization.}

During the reionization epoch, the characteristic bubble size (defined
here as the spherically averaged mean radius of the \HII regions that
contain most of the ionized volume\cite{FZH1}) increased with time as
smaller bubbles combined until their overlap completed and the diffuse
IGM was reionized.  However the largest size of isolated bubbles
(fully surrounded by \HI boundaries) that can be {\em observed} is
finite, because of the combined phenomena of cosmic variance and
causality. Figure 1 presents a schematic illustration of the
geometry. There is a surface on the sky corresponding to the time
along different lines-of-sight when the diffuse (uncollapsed) IGM was
{\em most recently neutral}. We refer to it as the Surface of Bubble
Overlap (SBO). There are two competing sources for fluctuations in the
SBO, each of which is dependent on the characteristic size, $R_{\rm
SBO}$, of the ionized regions just before the final overlap. First,
the finite speed of light implies that photons observed from different
points along the curved boundary of an \HII region must have been
emitted at different times during the history of the universe.
Second, bubbles on a co-moving scale $R$ achieve reionization over a
spread of redshifts due to cosmic variance in the initial conditions
of the density field smoothed on that scale.  The characteristic scale
of \HII bubbles grows with time, leading to a decline in the spread of
their formation redshifts\cite{BL1} as the cosmic variance is averaged
over an increasing spatial volume.  However the light-travel time
across a bubble rises concurrently. Suppose a signal photon which
encodes the presence of neutral gas (e.g. a 21cm line photon), is emitted
from the far edge of the ionizing bubble. If the adjacent region along
the line-of-sight has not become ionized by the time this photon
reaches the near side of the bubble, then the photon will encounter
diffuse neutral gas. Other photons emitted at this lower redshift will
therefore also encode the presence of diffuse neutral gas, implying
that the first photon was emitted prior to overlap, and not from the
SBO. Hence the largest observable scale of \HII regions when their
overlap completes, corresponds to the first epoch at which the light
crossing time becomes larger than the spread in formation times of
ionized regions.  Only then will the signal photon leaving the far
side of the HII region have the lowest redshift of any signal photon
along that line-of-sight.

The observed spectra of all quasars beyond $z\sim6.1$ each show a
Gunn-Peterson trough\cite{GP}$^,$\cite{f2}, a blank spectral region at
wavelengths shorter than \lya at the quasar redshift, indicating the
presence of \HI in the diffuse IGM.  The detection of Gunn-Peterson
troughs indicates a rapid change\cite{f1}$^,$\cite{Pe}$^,$\cite{WB1}
in the neutral content of the IGM at $z\sim6$, and hence a rapid
change in the intensity of the background ionizing flux. This rapid
change implies that overlap, and hence the reionization epoch,
concluded near $z\sim6$. The most promising observational
probe\cite{ZFH}$^,$\cite{MH1} of the reionization epoch is redshifted
21cm emission from intergalactic \HI.  Future observations using low
frequency radio arrays (e.g. the Low Frequency Array, LOFAR) will
allow a direct determination of the topology and duration of the phase
of bubble overlap.  Here we determine the expected angular scale and
redshift width of the 21cm fluctuations at the SBO theoretically, and
show that our determination is consistent with current observational
constraints.

We start by quantifying the constraints of causality and cosmic
variance. First suppose we have an \HII region with a physical radius
$R/(1+\langle z\rangle)$. The light crossing time of this radius is
\begin{equation}
\label{causality}
\langle\Delta z^2\rangle^{1/2} =
\left|\frac{dz}{dt}\right|_{\langle z\rangle}
\frac{R}{c(1+\langle z\rangle)},
\end{equation}
where at the high-redshifts of interest
$(dz/dt)=-(H_0\sqrt{\Omega_m})(1+z)^{5/2}$.  Here, $c$ is the speed of
light, $H_0$ is the present-day Hubble constant, $\Omega_m$ is the present
day matter density parameter, and $\langle z\rangle$ is the mean redshift
of the SBO.  Note that when discussing this crossing time, we are referring
to photons used to probe the ionized bubble (e.g. at 21cm), rather than
photons involved in the dynamics of the bubble evolution.

Second, overlap would have occurred at different times in different regions
of the IGM due to the cosmic scatter in the process of structure formation
within finite spatial volumes\cite{BL1}. Reionization should be completed
within a region of co-moving radius $R$ when the fraction of mass
incorporated into collapsed objects in this region attains a certain
critical value, corresponding to a threshold number of ionizing photons
emitted per baryon. The ionization state of a region is governed by the
enclosed ionizing luminosity, by its over-density, and by dense pockets of
neutral gas that are self shielding to ionizing radiation.  There is an
offset\cite{BL1} $\delta z$ between the redshift when a region of mean
over-density $\bar{\delta}_{\rm R}$ achieves this critical collapsed
fraction, and the redshift ${\bar z}$ when the universe achieves the same
collapsed fraction on average.  This offset may be computed\cite{BL1} from
the expression for the collapsed fraction\cite{BCE1} $F_{\rm col}$ within a
region of over-density $\bar{\delta}_{\rm R}$ on a co-moving scale $R$,
\begin{equation}
\label{scatter}
F_{\rm col}(M_{\rm min})=\mbox{erfc}\left[\frac{\delta_{\rm
c}-\bar{\delta}_{\rm R}}{\sqrt{2[\sigma_{\rm R_{\rm min}}^2-\sigma_{\rm
R}^2]}}\right],\hspace{5mm}\mbox{yielding}\hspace{5mm} \frac{\delta
z}{(1+\bar{z})}=\frac{\bar{\delta}_{\rm R}}{\delta_{\rm
c}(\bar{z})}-\left[1-\sqrt{1-\frac{\sigma_{\rm R}^2}{\sigma_{\rm R_{\rm
min}}^2}}\right],
\end{equation}
where $\delta_{\rm c}(\bar{z})\propto (1+\bar{z})$ is the collapse
threshold for an over-density at a redshift $\bar{z}$; $\sigma_{\rm R}$ and
$\sigma_{R_{\rm min}}$ are the variances in the power-spectrum linearly
extrapolated to $z=0$ on co-moving scales corresponding to the region of
interest and to the minimum galaxy mass $M_{\rm min}$, respectively.  The
offset in the ionization redshift of a region depends on its linear
over-density, $\bar{\delta}_{\rm R}$. As a result, the distribution of
offsets, and therefore the scatter in the SBO may be obtained directly from
the power spectrum of primordial inhomogeneities. As can be seen from
equation~(\ref{scatter}), larger regions have a smaller scatter due to
their smaller cosmic variance.

Note that equation~(\ref{scatter}) is independent of the critical
value of the collapsed fraction required for reionization. Moreover,
our numerical constraints are very weakly dependent on the minimum
galaxy mass, which we choose to have a virial temperature of $10^4$K
corresponding to the cooling threshold of primordial atomic gas.  The
growth of an \HII bubble around a cluster of sources requires that the
mean-free-path of ionizing photons be of order the bubble radius or
larger. Since ionizing photons can be absorbed by dense pockets of
neutral gas inside the \HII region, the necessary increase in the
mean-free-path with time implies that the critical collapsed fraction
required to ionize a region of size $R$ increases as well. This larger
collapsed fraction affects the redshift at which the region becomes
ionized, but not the scatter in redshifts from place to place which is
the focus of this {\em Letter}. Our results are therefore independent
of assumptions about unknown quantities such as the star formation
efficiency and the escape fraction of ionizing photons from galaxies,
as well as unknown processes of feedback in galaxies and clumping of
the IGM.

Figure~2 displays our two fundamental constraints.  The causality
constraint (Eq.~\ref{causality}) is shown as the blue line, giving a
longer crossing time for a larger bubble size. This contrasts with the
constraint of cosmic variance (Eq.~\ref{scatter}), indicated by the
red line, which shows how the scatter in formation times decreases
with increasing bubble size. The scatter in the SBO redshift and the
corresponding fluctuation scale of the SBO are given by the
intersection of these curves. We find that the thickness of the SBO is
$\langle\Delta z^2\rangle^{1/2}\sim0.13$, and that the bubbles which
form the SBO have a characteristic co-moving size of $\sim60$Mpc
(equivalent to 8.6 physical Mpc). At $z\sim6$ this size corresponds to
angular scales of $\theta_{\rm SBO}\sim0.4$ degrees on the sky.  We
have also examined a second model for the formation of \HII regions,
where ionization is reached when the collapsed fraction divided by the
density contrast exceeds a critical value. This model allows for the
possibility that the increased recombination rate offsets reionization
in overdense regions.
The increased range of formation times in this case leads
to a slightly larger value for the scale of overlapping \HII regions,
and we find $\langle\Delta z^2\rangle^{1/2}\sim0.2$, $R_{\rm
SBO}\sim90$Mpc and $\theta_{\rm SBO}\sim0.6$ degrees.

A scatter of $\sim0.15$ in the SBO is somewhat larger than the value
extracted from existing numerical
simulations\cite{g1}$^,$\cite{y1}. The difference is most likely due
to the limited size of the simulated volumes; while the simulations
appropriately describe the reionization process within limited regions
of the universe, they are not sufficiently large to describe the
global properties of the overlap phase\cite{BL1}. The scales over
which cosmological radiative transfer has been simulated are smaller
than the characteristic extent of the SBO, which we find to be $R_{\rm
SBO}\sim70$ co-moving Mpc.

We can constrain the scatter in the SBO redshift observationally using
the spectra of the highest redshift quasars. Since only a trace amount
of neutral hydrogen is needed to absorb Ly$\alpha$ photons, the time
where the IGM becomes \lya transparent need not coincide with bubble
overlap. Following overlap the IGM was exposed to ionizing sources in
all directions and the ionizing intensity rose rapidly. After some
time the ionizing background flux was sufficiently high that the \HI
fraction fell to a level at which the IGM allowed transmission of
resonant \lya photons. This is shown schematically in Figure~1.  The
lower wavelength limit of the Gunn-Peterson trough corresponds to the
\lya wavelength at the redshift when the IGM started to allow
transmission of \lya photons {\em along that particular
line-of-sight}.  In addition to the SBO we therefore also define the
Surface of \lya Transmission (hereafter SLT) as the redshift along
different lines-of-sight when the diffuse IGM became transparent to
Ly$\alpha$ photons.

The scatter in the SLT redshift is an observable which we would like to
compare with the scatter in the SBO redshift.  The variance of the density
field on large scales results in the biased clustering of
sources\cite{BL1}.  \HII regions grow in size around these clusters of
sources. In order for the ionizing photons produced by a cluster to advance
the walls of the ionized bubble around it, the mean-free-path of these
photons must be of order the bubble size or larger.  After bubble overlap,
the ionizing intensity at any point grows until the ionizing photons have
time to travel across the scale of the new mean-free-path, which represents
the horizon out to which ionizing sources are visible.  Since the
mean-free-path is larger than $R_{\rm SBO}$, the ionizing intensity at the
SLT averages the cosmic scatter over a larger volume than at the SBO.  This
constraint implies that the cosmic variance in the SLT redshift must be
smaller than the scatter in the SBO redshift. However, it is possible that
opacity from small-scale structure contributes additional scatter to the
SLT redshift.

If cosmic variance dominates the observed scatter in the SLT redshift, then
based on the spectra of the three $z>6.1$ quasars\cite{f2}$^,$\cite{WB1} we
would expect the scatter in the SBO redshift to satisfy $\langle\Delta
z^2\rangle^{1/2}_{\rm obs}\ga0.05$. In addition, analysis of the {\it
proximity effect} for the size of the \HII regions around the two highest
redshift quasars\cite{WL1}$^,$\cite{mh1} implies a neutral fraction that is
of order unity (i.e. pre-overlap) at $z\sim6.2-6.3$, while the transmission
of Ly$\alpha$ photons at $z\la6$ implies that overlap must have completed
by that time. This restricts the scatter in the SBO to be $\langle\Delta
z^2\rangle^{1/2}_{\rm obs}\la0.25$. The constraints on values for the
scatter in the SBO redshift are shaded gray in Figure~\ref{fig2}.  It is
reassuring that the theoretical prediction for the SBO scatter of
$\langle\Delta z^2\rangle^{1/2}_{\rm obs}\sim0.15$, with a characteristic
scale of $\sim70$ co-moving Mpc, is bounded by these constraints.

The presence of a significantly neutral IGM just beyond the redshift
of overlap\cite{WL1}$^,$\cite{mh1} is encouraging for upcoming 21cm
studies of the reionization epoch as it results in emission near an
observed frequency of 200 MHz where the signal is most readily
detectable. Future observations of redshifted 21cm line emission at
$6\la z\la 6.5$ with instruments such as the Low Frequency Array
(LOFAR), will be able to map the three-dimensional distribution of HI
at the end of reionization. The intergalactic \HII regions will
imprint a 'knee' in the power-spectrum of the 21cm anisotropies on a
characteristic angular scale corresponding to a typical isolated \HII
region\cite{ZFH}. Our results suggest that this characteristic angular
scale is large at the end of reionization, $\theta_{\rm SBO}\sim
0.5$ degrees, motivating the construction of compact low
frequency arrays. An SBO thickness of $\langle \Delta
z^2\rangle^{1/2}\sim0.15$ suggests a minimum frequency
band-width of $\sim8$ MHz for experiments aiming to detect
anisotropies in 21cm emission just prior to overlap. These results
will help guide the design of the next generation of low-frequency
radio observatories in the search for 21cm emission at the end of the
reionization epoch.


\vskip 0.2in
\noindent
{ACKNOWLEDGMENTS.} This work was supported in part by grants from ARC,
NSF and NASA.

\vskip 1in

\begin{figure*}[htbp]
\epsscale{1.}  
\plotone{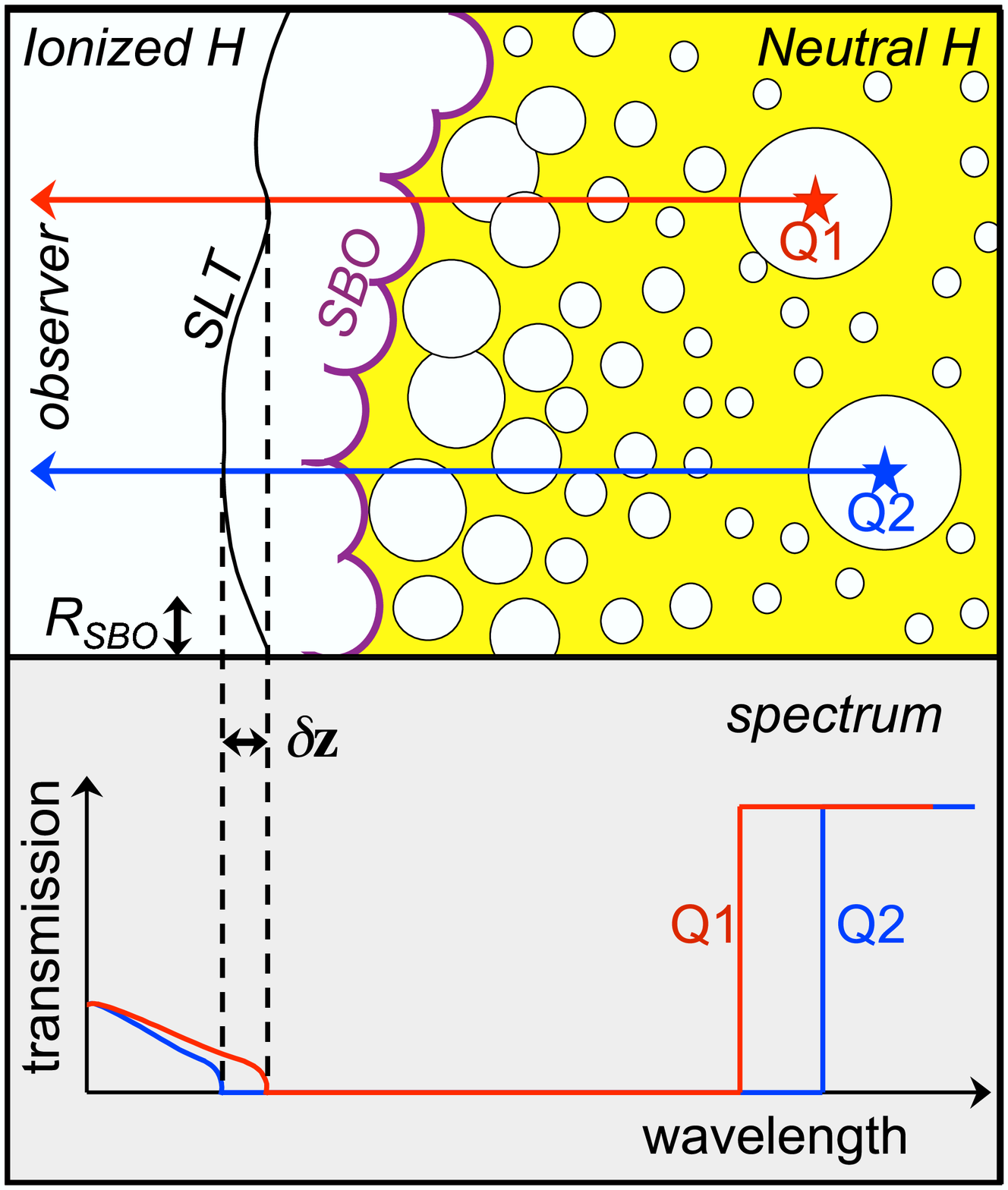}
\end{figure*}

\begin{figure*}[htbp]
\caption{\label{fig1} The distances to the observed Surface of Bubble
Overlap (SBO) and Surface of Ly$\alpha$ Transmission (SLT) fluctuate
on the sky. The SBO corresponds to the first region of diffuse neutral
IGM {\em observed} along a random line-of-sight. It fluctuates across
a shell with a minimum width dictated by the condition that the light
crossing time across the characteristic radius $R_{\rm SBO}$ of
ionized bubbles equals the cosmic scatter in their formation
times. Thus, {\it causality} and {\it cosmic variance} determine the
characteristic scale of bubbles at the completion of bubble overlap.
After some time delay the IGM becomes transparent to Ly$\alpha$
photons, resulting in a second surface, the SLT.  The upper panel
illustrates how the lines-of-sight towards two quasars (Q1 in red and
Q2 in blue) intersect the SLT with a redshift difference $\delta
z$. The resulting variation in the observed spectrum of the two
quasars is shown in the lower panel.  Observationally, the ensemble of
redshifts down to which the Gunn-Peterson troughs are seen in the
spectra of $z>6.1$ quasars is drawn from the probability distribution
$dP/dz_{\rm SLT}$ for the redshift at which the IGM started to allow
\lya transmission along random lines-of-sight. The observed values of
$z_{\rm SLT}$ show a small scatter\cite{f2} in the SLT redshift around
an average value of $\langle z_{\rm SLT}\rangle\approx 5.95$. Some
regions of the IGM may have also become transparent to Ly$\alpha$
photons prior to overlap, resulting in windows of transmission inside
the Gunn-Peterson trough (one such region may have been seen\cite{WB1}
in SDSS J1148+5251).  In the existing examples, the portions of the
universe probed by the lower end of the Gunn-Peterson trough are
located several hundred co-moving Mpc away from the background quasar,
and are therefore not correlated with the quasar host galaxy. The
distribution $dP/dz_{\rm SLT}$ is also independent of the redshift
distribution of the quasars. Moreover, lines-of-sight to these quasars
are not causally connected at $z\sim 6$ and may be considered
independent. }
\end{figure*}
\clearpage

\begin{figure*}[htbp]
\epsscale{1.}  
\plotone{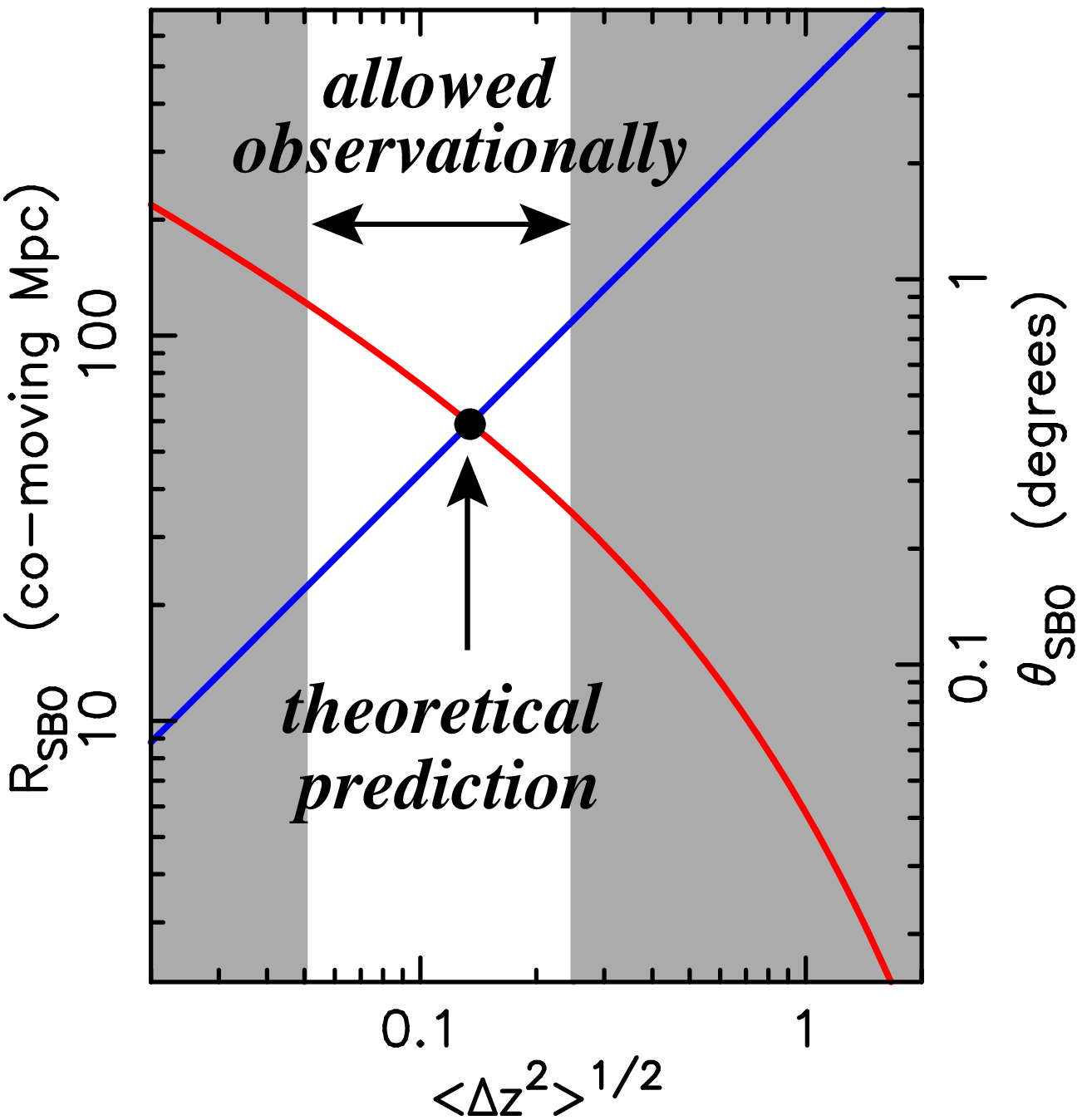}
\end{figure*}

\begin{figure*}[htbp]
\caption{\label{fig2} Constraints on the scatter in the SBO redshift and
the characteristic size of isolated bubbles at the final overlap stage,
$R_{\rm SBO}$ (see Fig. 1). The characteristic size of \HII regions grows
with time. The SBO is observed for the bubble scale at which the light
crossing time (blue line) first becomes smaller than the cosmic scatter in
bubble formation times (red line).  At $z\sim6$, the implied scale $R_{\rm
SBO}\sim60$ co-moving Mpc (or $\sim 8.6$ physical Mpc), corresponds to a
characteristic angular radius of $\theta_{\rm SLT}\sim 0.4$ degrees on the
sky.  After bubble overlap, the ionizing intensity grows to a level at
which the IGM becomes transparent to Ly$\alpha$ photons. The collapsed
fraction required for \lya transmission within a region of a certain size
will be larger than required for its ionization. However, the scatter in
equation~(\ref{scatter}) is not sensitive to the collapsed fraction, and so
may be used for both the SBO and SLT.  The scatter in the SLT is smaller
than the cosmic scatter in the structure formation time on the scale of the
mean-free-path for ionizing photons.  This mean-free-path must be longer
than $R_{\rm SBO}\sim60$Mpc, an inference which is supported by analysis of
the Ly$\alpha$ forest at $z\sim4$ where the mean-free-path is
estimated\cite{ME1} to be $\sim 120$ co-moving Mpc at the Lyman limit (and
longer at higher frequencies). If it is dominated by cosmic variance, then
the scatter in the SLT redshift provides a lower limit to the SBO
scatter. The three known quasars at $z>6.1$ have \lya transmission
redshifts of\cite{WB1}$^,$\cite{f2} $z_{\rm SLT}=5.9$, 5.95 and 5.98,
implying that the scatter in the SBO must be $\ga0.05$ (this scatter may
become better known from follow-up spectroscopy of Gamma Ray Burst
afterglows at $z>6$ that might be discovered by the {\it SWIFT}
satellite\cite{BL2}$^,$\cite{BL3}).  
The observed scatter in the SLT redshift is somewhat smaller than the
predicted SBO scatter, confirming the expectation that cosmic variance is
smaller at the SLT. The scatter in the SBO redshift must also be $\la0.25$
because the lines-of-sight to the two highest redshift quasars have a
redshift of \lya transparency at $z\sim6$, but a neutral fraction that is
known from the {\it proximity effect}\cite{WL1} to be substantial at
$z\ga6.2-6.3$. The excluded regions of scatter for the SBO are shown in
gray.  Throughout this {\it Letter}, we adopt the latest values for the
cosmological parameters as inferred from the {\em Wilkinson Microwave
Anisotropy Probe} data\cite{SB1}.}
\end{figure*}


\small
\noindent
\begin{thebibliography}{}




\v
\bibitem[$^{1}$]{BL1}\cite{BL1} Barkana, R., Loeb, A., Unusually large 
fluctuations in the statistics of galaxy formation at high redshift,
{\it Astrophys. J.}, {\bf 609}, 474-481, (2004)

\v
\bibitem[$^{2}$]{FZH1}\cite{FZH1} Furlanetto, S.R., Zaldarriaga, M., 
Hernquist, L., The growth of \HII regions during reionization, {\it
Astrophys. J.}, submitted, astro-ph/0403697

\v
\bibitem[$^{3}$]{GP}\cite{GP} Gunn, J. E., Peterson, B. A., On the
density of neutral hydrogen in intergalactic space, {\it Astrophys. J.}
{\bf 142}, 1633-1641 (1965)

\v
\bibitem[$^{4}$]{f2}\cite{f2} Fan, X., \textit{et al.}, A survey of $z>5.7$ quasars in the Sloan Digital Sky Survey III: discovery of five additional quasars, {\it Astron. J.}, in press, astro-ph/0405138

\v
\bibitem[$^{5}$]{f1}\cite{f1} Fan, X., \textit{et al.}, Evolution of the
ionizing background and the epoch of reionization from the spectra of
$z\sim 6$ quasars, {\it Astron. J.} {\bf 123}, 1247-1257 (2002)

\v
\bibitem[$^{6}$]{Pe}\cite{Pe} Pentericci, L.~et al., VLT optical and
near-infrared observations of the $z=6.28$ quasar SDSS J1030+0524, {\it
Astron. J.}, {\bf 123}, 2151-2158 (2002)

\v
\bibitem[$^{7}$]{WB1}\cite{WB1} White, R.L., Becker, R.H., Fan, X.,
Strauss, M.A., Probing the ionization state of the universe at $z>6$,
{\it Astron. J.} {\bf 126}, 1-14, (2003)

\v
\bibitem[$^{8}$]{ZFH}\cite{ZFH} Zaldarriaga, M., Furlanetto, S.R.,
Hernquist, L., 21 Centimeter Fluctuations from Cosmic Gas at High
Redshifts, {\it Astrophys. J.}, submitted, astro-ph/0311514

\v
\bibitem[$^{9}$]{MH1}\cite{MH1} 
Morales, M.F., Hewitt, J., Toward Epoch of Reionization Measurements with Wide-Field Radio Observations, {\it Astrophys. J.}, accepted, astro-ph/0312437

\v
\bibitem[$^{10}$]{BCE1}\cite{BCE1} 
Bond, J. R.; Cole, S.; Efstathiou, G.; Kaiser, N., Excursion set mass functions for hierarchical Gaussian fluctuations, {\it Astrophys. J.}, {\bf 379}, 440-460 (1991)

\v
\bibitem[$^{11}$]{g1}\cite{g1} Gnedin, N.Y., Cosmological reionization by stellar sources, {\it Astrophys. J.}, {\bf 535}, 530-554, (2000)

\v
\bibitem[$^{12}$]{y1}\cite{y1} Yoshida, N., Sokasian, A., Hernquist, L., Springel, V., Early structure formation and reionization in a cosmological model with a running primordial power spectrum, {\it Astrophys. J.}, {\bf 598}, 73-85, (2003)

\v
\bibitem[$^{13}$]{WL1}\cite{WL1} Wyithe, J.S.B., Loeb, A., A large neutral fraction of cosmic hydrogen a billion years after the Big Bang, {\it Nature} {\bf 427}, 815-817, (2004)

\v
\bibitem[$^{14}$]{mh1}\cite{mh1} Mesinger, A., Haiman, Z., Evidence for a cosmological Stromgren surface and for significant neutral hydrogen surrounding the quasar SDSS J1030+0524, {\it Astrophys. J. Lett.}, in press, astro-ph/0406188

\v
\bibitem[$^{15}$]{ME1}\cite{ME1} Miralda-Escude, J., On the Evolution of
the Ionizing Emissivity of Galaxies and Quasars Required by the Hydrogen
Reionization, {\it Astrophys. J.}, {\bf 597}, 66-73, (2003)

\v
\bibitem[$^{16}$]{BL2}\cite{BL2} Barkana, R., Loeb, A., GRBs verses
quasars: Lyman-$\alpha$ signatures of reionization verses cosmological
infall, {\it Astrophys. J.}, {\bf 601}, 64-77 (2004)

\v
\bibitem[$^{17}$]{BL3}\cite{BL3} Bromm, V., Loeb, A., The expected 
redshift distribution of Gamma-Ray Bursts, {\it Astrophys. J.}, {\bf
575}, 111-116, (2002)

\v
\bibitem[$^{18}$]{SB1}\cite{SB1} Spergel, D. N, et al., First-year
Wilkinson microwave anisotropy probe (WMAP) observations:
determination of cosmological parameters, {\it Astrophys. J. Supp.}, {\bf
148}, 175-194 (2003)

\end{thebibliography}



\normalsize
\end{document}